\documentclass[10pt,aps,prl,twocolumn,superscriptaddress]{revtex4-1}

\usepackage{natbib}
\usepackage[dvips]{graphicx}
\usepackage{bm}
\usepackage{amsmath, amssymb}

\newcommand{\BFA}{BaFe$_{2}$As$_{2}$}
\newcommand{\BFCA}{Ba(Fe$_{1-x}$Co$_{x}$)$_{2}$As$_{2}$}

\newcommand{\TN}{$T_{\mathrm{N}}$}
\newcommand{\Ts}{$T_{\mathrm{s}}$}
\newcommand{\Tc}{$T_{\mathrm{c}}$}
\newcommand{\cm}{cm$^{-1}$}

\newcommand{\sa}{$\sigma_a$}
\newcommand{\sib}{$\sigma_b$}
\newcommand{\saw}{$\sigma_a(\omega)$}
\newcommand{\sbw}{$\sigma_b(\omega)$}

\begin{document}

\preprint{ver.\ 7}

\title{Effect of Co doping on the in-plane anisotropy in the optical spectrum of underdoped \BFCA{}}



\author{M.~Nakajima}
\email[]{m-nakajima@aist.go.jp}
\affiliation{Department of Physics, University of Tokyo, Tokyo 113-0033, Japan}
\affiliation{National Institute of Advanced Industrial Science and Technology, Tsukuba 305-8568, Japan}
\affiliation{JST, Transformative Research-Project on Iron Pnictides (TRIP), Tokyo 102-0075, Japan}
\author{S.~Ishida}
\affiliation{Department of Physics, University of Tokyo, Tokyo 113-0033, Japan}
\affiliation{National Institute of Advanced Industrial Science and Technology, Tsukuba 305-8568, Japan}
\affiliation{JST, Transformative Research-Project on Iron Pnictides (TRIP), Tokyo 102-0075, Japan}
\author{Y.~Tomioka}
\affiliation{National Institute of Advanced Industrial Science and Technology, Tsukuba 305-8568, Japan}
\affiliation{JST, Transformative Research-Project on Iron Pnictides (TRIP), Tokyo 102-0075, Japan}
\author{K.~Kihou}
\affiliation{National Institute of Advanced Industrial Science and Technology, Tsukuba 305-8568, Japan}
\affiliation{JST, Transformative Research-Project on Iron Pnictides (TRIP), Tokyo 102-0075, Japan}
\author{C.~H.~Lee}
\affiliation{National Institute of Advanced Industrial Science and Technology, Tsukuba 305-8568, Japan}
\affiliation{JST, Transformative Research-Project on Iron Pnictides (TRIP), Tokyo 102-0075, Japan}
\author{A.~Iyo}
\affiliation{National Institute of Advanced Industrial Science and Technology, Tsukuba 305-8568, Japan}
\affiliation{JST, Transformative Research-Project on Iron Pnictides (TRIP), Tokyo 102-0075, Japan}
\author{T.~Ito}
\affiliation{National Institute of Advanced Industrial Science and Technology, Tsukuba 305-8568, Japan}
\affiliation{JST, Transformative Research-Project on Iron Pnictides (TRIP), Tokyo 102-0075, Japan}
\author{T.~Kakeshita}
\affiliation{Department of Physics, University of Tokyo, Tokyo 113-0033, Japan}
\affiliation{JST, Transformative Research-Project on Iron Pnictides (TRIP), Tokyo 102-0075, Japan}
\author{H.~Eisaki}
\affiliation{National Institute of Advanced Industrial Science and Technology, Tsukuba 305-8568, Japan}
\affiliation{JST, Transformative Research-Project on Iron Pnictides (TRIP), Tokyo 102-0075, Japan}
\author{S.~Uchida}
\affiliation{Department of Physics, University of Tokyo, Tokyo 113-0033, Japan}
\affiliation{JST, Transformative Research-Project on Iron Pnictides (TRIP), Tokyo 102-0075, Japan}


\date{\today}

\begin{abstract}

We investigated the anisotropy in the in-plane optical spectra of detwinned \BFCA{}. The optical conductivity spectrum of \BFA{} shows appreciable anisotropy in the magnetostructural ordered phase, whereas the dc ($\omega=0$) resistivity is almost isotropic at low temperatures. Upon Co doping, the resistivity becomes highly anisotropic, while the finite-energy \textit{intrinsic} anisotropy is suppressed. It is found that anisotropy in resistivity arises from anisotropic impurity scattering from doped Co atoms, \textit{extrinsic} in origin. Intensity of a specific optical phonon mode is also found to show striking anisotropy in the ordered phase. The anisotropy induced by Co impurity and that observed in the optical phonon mode are hallmarks of the highly polarizable electronic state in the ordered phase.

\end{abstract}


\maketitle

A parent compound of iron-arsenide superconductors, \textit{e.g.}, \BFA{}, exhibits a magnetostructural phase transition from a paramagnetic-tetragonal (PT) to an antiferromagnetic-orthorhombic (AFO) phase. The antiferromagnetic order is of stripe type --- Fe spins align antiferromagnetically along the longer crystallographic $a$ axis and ferromagnetically along the shorter $b$ axis \cite{Cruz2008,Rotter2008,Huang2008}. An anisotropic electronic state realized in the AFO phase of undoped and underdoped iron arsenides has attracted much interest as a proximate phase to superconductivity \cite{Fisher2011}. The resistivity measurement on detwinned single crystals of \BFCA{} revealed that the resistivity along the $b$ axis is higher than that along the $a$ axis \cite{Chu2010,Tanatar2010}. Spectroscopic experiments, such as neutron scattering \cite{Zhao2009}, scanning tunneling spectroscopy \cite{Chuang2010}, and angle-resolved photoemission spectroscopy (ARPES) \cite{Shimojima2010,Yi2011}, were also successful in observing the electronic anisotropy in the iron arsenides.

The origin of the anisotropy in the AFO phase which breaks the fourfold symmetry of the PT phase has been discussed in terms of either anisotropic spin exchange interactions \cite{Fang2008,Xu2008,Fernandes2010} or orbital ordering/polarization with inequivalent Fe 3d$_{xz}$ and 3d$_{yz}$ orbitals \cite{Lee2009,Kruger2009,Yin2010,Daghofer2010,Chen2010,Valenzuela2010}. In this context, the anisotropic resistivity in the parent compound has often been interpreted as related to the inherently anisotropic electronic structure of the AFO phase \cite{Tanatar2010,Valenzuela2010,Yin2011,Sugimoto2011}. However, it has been shown later that an annealing process makes the resistivity of \BFA{} much less anisotropic \cite{Nakajima2011,Ishida2011}, and robust anisotropy shows up in the optical spectrum \cite{Nakajima2011}.

With Co doping, the transition from the AFO to the PT phase is suppressed. Despite the suppression of the AFO order, the anisotropy in the in-plane resistivity is enhanced. Therefore, the origin of the resistivity anisotropy is not trivial, and the effect of the Co doping should be re-examined from a viewpoint different from simply carrier doping.

\begin{figure}[b]
\includegraphics[width=80mm,clip]{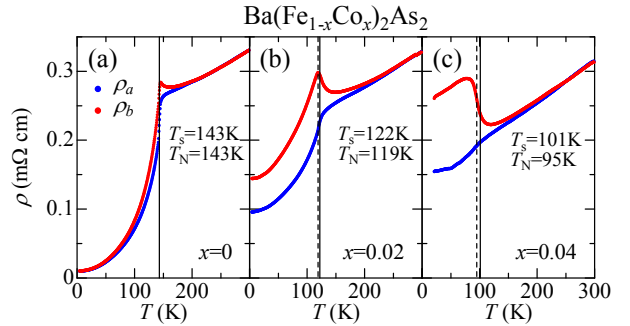}%
\caption{(Color online) Temperature dependence of $\rho_a$ and $\rho_b$ for annealed \BFCA{} with (a) $x=0$, (b) 0.02, and (c) 0.04. Solid and broken vertical lines indicate \Ts{} and \TN{}, respectively, determined from the resistivity of a twinned crystal \cite{Chu2010,Ishida}.}
\end{figure}

\begin{figure*}
\includegraphics[width=150mm,clip]{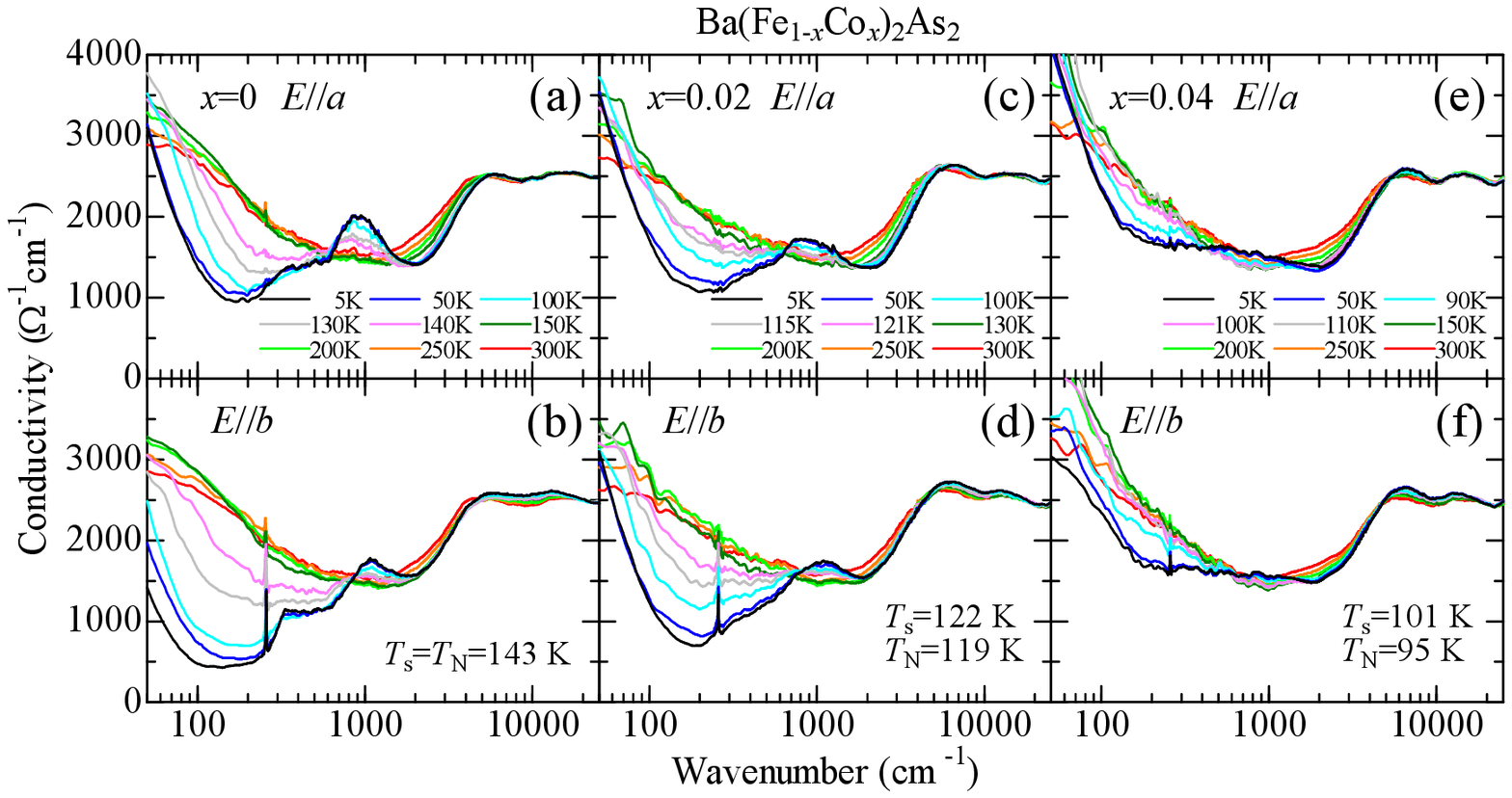}%
\caption{(Color online) Temperature evolution of \saw{} and \sbw{} of detwinned \BFCA{} ($x=0$, 0.02, and 0.04). The crystal $b$ axis corresponds to the direction of applied compressive pressure. \sa{} (\sib{}) of undoped \BFA{} in the AFO phase is reduced below 600 (850) \cm{}, and, by piling up the reduced spectral weight, a peak develops at 950 (1050) \cm{}. For $x=0.02$ and 0.04, the low-energy conductivity below 500 (700) \cm{} and 400 (500) \cm{} is suppressed, forming a peak at 900 (950) \cm{} and 750 (800) \cm{}, respectively.}
\end{figure*}

In this letter, we present the optical spectrum measured on detwinned single crystals of underdoped \BFCA{} ($x=0$, 0.02, and 0.04). As previously reported, remarkable anisotropy is observed in the optical conductivity spectrum of \BFA{} persisting up to 17000 \cm{} ($\sim$ 2 eV), while anisotropy in the dc ($\omega=0$) resistivity is hidden by an isotropic Drude component at low temperatures \cite{Nakajima2011}. We show that, whereas the anisotropy in the high-energy optical conductivity decreases with Co doping, it shows up in the width of the Drude component, which increases as doping proceeds. This gives evidence that a doped Co atom forms an impurity state which anisotropically scatters carriers. We also observe striking anisotropy in the intensity of a particular optical phonon mode at 257 \cm{} in the AFO phase. The phonon anisotropy weakens with Co doping, associated with the suppression of the AFO order.


Single crystals of \BFCA{} ($x=0$, 0.02, and 0.04) were grown and post-annealed according to the method described in Refs.\ \onlinecite{Nakajima2010,Nakajima2011}. The structural (magnetic) transition temperatures, \Ts{} (\TN{}), determined from the resistivity curve of free-standing twinned samples are 143 (143) K, 122 (119) K, and 101 (95) K for $x=0$, 0.02, and 0.04, respectively \cite{Chu2010,Ishida}. To remove twinned domains, we applied uniaxial compressive pressure of approximately 50 MPa. The anisotropy in resistivity becomes maximal at this pressure and ceases to increase with further application of pressure. Optical reflectivity was measured with the $a$- and $b$-axis polarized light in the energy range of 50--40000 \cm{} using a Fourier-transform-type infrared spectrometer (Bruker IFS113v) and a grating monochromator (JASCO CT-25C). The optical conductivity was derived from the Kramers-Kronig transformation of the reflectivity spectrum. In the higher-energy region, we used the spectrum of \BFA{} measured using vacuum ultraviolet synchrotron radiation (Supporting Information in Ref.\ \onlinecite{Nakajima2011}). The Hagen-Rubens and/or Drude-Lorentz formula was used for the low-energy extrapolation in order to smoothly connect to the spectrum in the measured region and to fit the measured resistivity value at $\omega=0$.


Figures 1(a)-(c) show temperature ($T$) dependence of resistivity along the $a$ and $b$ axis ($\rho_a$ and $\rho_b$) for the three compositions of annealed \BFCA{}. Solid and broken vertical lines indicate \Ts{} and \TN{}, respectively. For all the compositions, near the transition temperature, $\rho_b$ jumps up, whereas $\rho_a$ drops rapidly, and then both decrease with lowering temperature in a similar manner. The anisotropy is seen at temperatures above \Ts{}. Note that, the resistivity of undoped \BFA{} is nearly isotropic at low temperatures. With Co doping, anisotropy shows up predominantly in the residual-resistivity (RR) component. The RR in either direction increases roughly proportional to the Co content $x$, indicating that the introduced Co atom anisotropically scatters carriers as an elastic scattering center.

Figures 2(a) and 2(b) display the $T$ evolution of the optical conductivity spectra of \BFA{} for the $a$- and $b$-axis direction (\saw{} and \sbw{}), respectively. Above \Ts{}, the conductivity shows a peak at $\omega=0$ and a long tail extending to a higher-energy region in which interband transitions dominate. No discernible anisotropy is observed between the $a$- and $b$-axis conductivity in the PT phase. Once temperature decreases below \Ts{} = 143 K, the spectrum exhibits a gap (pseudogap) feature associated with the development of the AFO order --- the low-energy conductivity is suppressed and transferred to a higher-energy region. The gap behavior is observed in both the $a$- and $b$-axis spectra, but the energy scale is different. The $b$-axis gap is larger than the $a$-axis one, and hence the low-energy conductivity is suppressed more deeply in the $b$ direction. The different gap-energy scale between the $a$ and $b$ direction is responsible for the anisotropy in the optical conductivity \cite{Nakajima2011}.

\begin{figure*}
\includegraphics[width=158mm,clip]{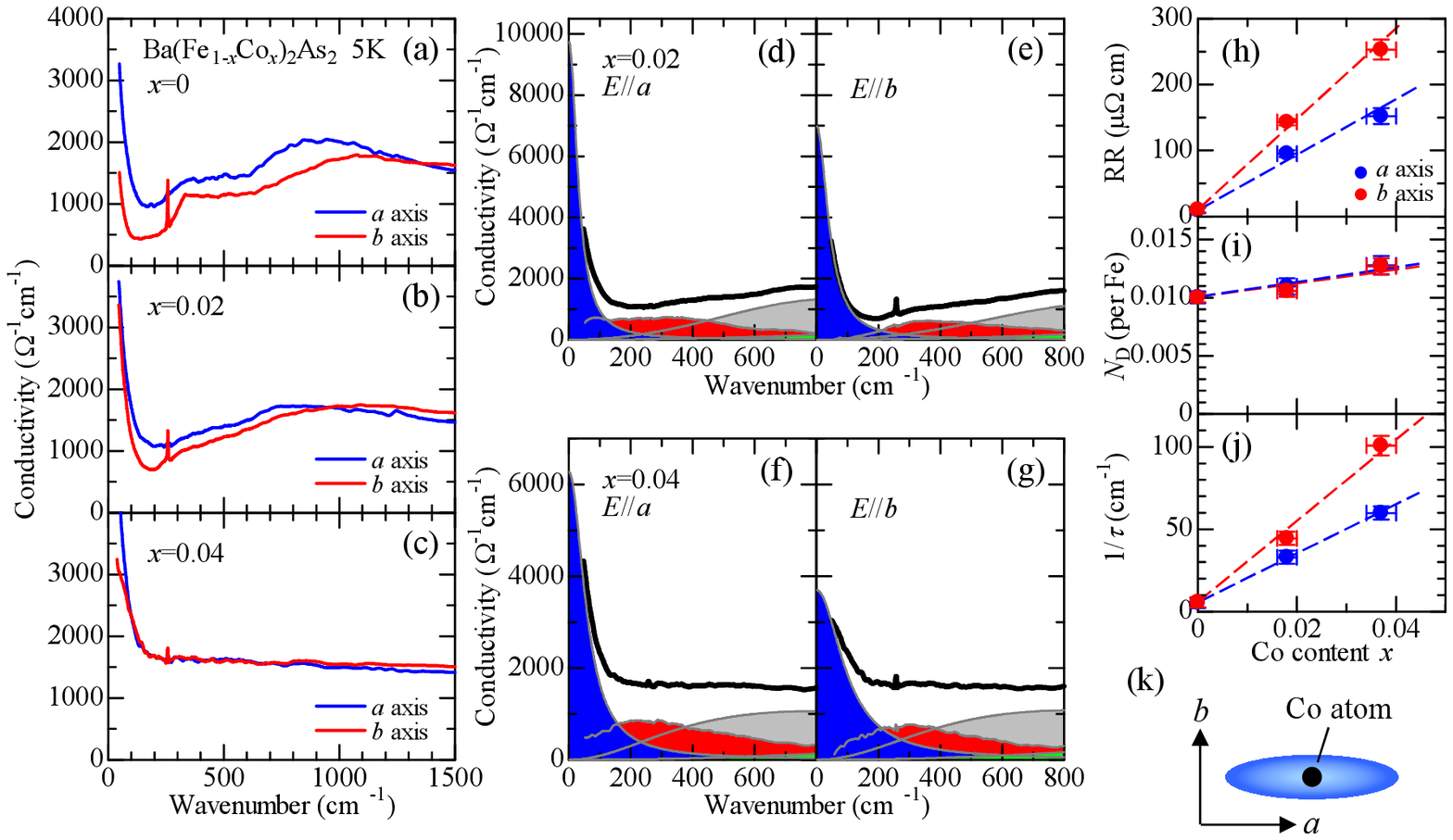}%
\caption{(Color online) Anisotropic optical conductivity spectra at $T=5$ K in the low-energy region below 1500 \cm{} for (a) $x=0$, (b) 0.02, and (c) 0.04. (d)--(g) Decomposition of \saw{} and \sbw{} for $x=0.02$ and 0.04. (h) Residual resistivity of $\rho_a$ and $\rho_b$ for \BFCA{} plotted against the Co content $x$. The broken lines are guides to the eye. (i) Drude weight ($N_{\mathrm{D}}$) and (j) width ($1/\tau$) extracted from the conductivity spectrum at 5 K plotted against $x$. For $x=0$, $1/\tau$ is $\sim$ 5 \cm{} for both directions \cite{Nakajima2011}. The broken lines are guides to the eye. (k) Schematic picture of an anisotropic impurity state around a doped Co atom.}
\end{figure*}

The $T$ evolution of the conductivity spectrum of the Co-doped compounds is similar to that of \BFA{} (Figs.\ 2(c)-(f)). As in the case of $x=0$, the gap feature becomes apparent only below \Ts{} (see the spectrum colored in pink). The Co doping makes the gap feature less clear and the gap energy scale smaller. Figures 3(a)-(c) show the anisotropy in the low-energy conductivity at $T=5$ K. For $x=0$, \sa{} is higher than \sib{} below 1350 \cm{}. Above 1350 \cm{}, the anisotropy is reversed and persists up to 17000 \cm{} ($\sim$ 2 eV) \cite{Nakajima2011}. Almost isotropic resistivity at low temperatures stems from the isotropic and sharp Drude component showing extremely high conductivity peak. With doping ($x=0.02$), the energy where the conductivity anisotropy is reversed decreases, and the magnitude of anisotropy in conductivity becomes smaller. For $x=0.04$, the anisotropy in the optical conductivity is considerably diminished. This is in sharp contrast with the dc resistivity which becomes more anisotropic with doping.



Here, we consider the origin of the resistivity anisotropy for the Co-doped compounds based on the optical conductivity spectrum. In the AFO phase, $\rho_a$ and $\rho_b$ show almost the same $T$ dependence, so the anisotropy in resistivity is attributed to the anisotropy in the $T$-independent residual component (Fig.\ 3(h)). Then, either the effective mass of carriers ($m^*$) and/or the carrier scattering time ($\tau$) appearing in the Drude formula, $\sigma=\frac{ne^2\tau}{m^*}$, should be anisotropic. An advantage of the optical spectrum is that one can estimate $n/m^*$ and $\tau$ separately by isolating a Drude component from the conductivity spectrum, the weight (area) and the width of which correspond to $n/m^*$ and $1/\tau$, respectively. We decompose the spectra for Co-doped compounds (Figs.\ 3(d)--(g)), as we did for \BFA{} in Ref.\ \onlinecite{Nakajima2011}. Since the gap feature, in particular the shoulder structure in \saw{} and the cusp one in \sbw{} at $\sim$ 350 \cm{} for \BFA{}, becomes smeared with Co doping, we fit the Drude and the higher-energy component (in blue and gray, respectively), and the rest is shown in red. Note that each component smoothly evolves with increasing $x$. The estimated Drude weight ($N_{\mathrm{D}} \propto n/m^*$) and the width ($1/\tau$) are plotted against the Co content $x$ in Figs.\ 3(i) and 3(j). $N_{\mathrm{D}}$ shows no significant anisotropy, and hence $m^*$ is nearly isotropic. This is consistent with the direct observation of almost isotropic Fermi-surface pockets by ARPES \cite{Yi2011} and quantum oscillation \cite{Terashima2011}. On the other hand, the Drude width becomes anisotropic upon Co doping, and the anisotropy in $1/\tau$ increases in proportion to the Co content. Therefore, the resistivity anisotropy originates from the anisotropy in the elastic carrier scattering from dopant Co atoms.

A question is what makes Co atoms anisotropic scattering centers. We speculate that a Co atom polarizes its electronic surroundings anisotropically, as schematically sketched in Fig.\ 3(k), and thereby forms an unique impurity potential with anisotropic scattering cross section. This speculation is supported by the result of scanning tunneling spectroscopy measurements \cite{Allan}. It is also suggested theoretically that local orbital order oriented in the $a$ axis is formed around an impurity \cite{Kontani2012}. 

\begin{figure}
\includegraphics[width=75mm,clip]{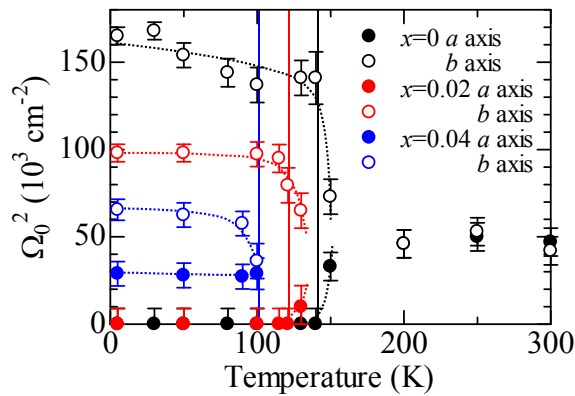}%
\caption{(Color online) Intensities (${\Omega_0}^2$) of the optical phonon mode at various temperatures for \BFCA{}. The solid vertical lines indicate \Ts{}'s for each composition.}
\end{figure}

For \BFA{}, the sharp peak centered at 257 \cm{} characterizes \sbw{} in the AFO phase. This peak is assigned to be an optical phonon mode corresponding to the in-plane relative displacements of Fe and As atoms \cite{Akrap2009,Schafgans2011}. In Fig.\ 4, we plot the intensity ${\Omega_0}^2$ of the phonon mode as a function of temperature. It is observed in both \saw{} and \sbw{} in the PT phase, but, in the AFO phase, its intensity is enhanced in \sbw{}, while reduced in \saw{} to an indiscernible level \cite{Nakajima2011}. Such striking anisotropy in the phonon intensity strongly suggests that this lattice vibration induces an additional polarization field in the AFO state. The induced polarization would almost cancel out the dipole field associated with this optical phonon in the $a$-axis direction, whereas it would enhance in the $b$-axis direction. Exactly the same behavior is seen in the spectra for $x=0.02$ (Figs.\ 2(c) and 2(d)), but the $b$-axis peak intensity is lower than that for the parent compound. For $x=0.04$, the $b$-axis phonon intensity further weakens, and the phonon mode is observed even in \saw{} in the AFO phase. The result indicates that the phonon-induced anisotropic electronic environment weakens associated with the weakening of the AFO order by Co doping.

The anisotropy of the optical phonon mode together with the anisotropic scattering by dopant Co atoms gives evidence of the unique anisotropic state in the AFO phase highly polarizable in response to static impurities and dynamical lattice distortions. It is highly probable that the stripe antiferromagnetic order and the orthorhombic lattice distortion alone are difficult to produce such anisotropy. In the present optical spectra of the Co-doped compounds, the gap feature becomes apparent at a temperature between \Ts{} and \TN{} (121 K and 100 K for $x=0.02$ and 0.04, respectively). More clearly, Dong \textit{et al.}\ reported the result for NdFeAsO (\Ts{}=150 K and \TN{}=140 K), indicating that the gap opens in the spectrum at 142 K above the AFO ordering temperature \TN{} \cite{Dong2010}. This gives support for substantial role played by orbital degree of freedom.

An anisotropic electronic state (nematicity) associated with pseudogap is one of the hottest issues for the quest of mechanism of high-\Tc{} superconductivity in cuprates \cite{Fradkin2010}. In this state, a fourfold symmetry in the underlying lattice is retained, but the electron system spontaneously forms an anisotropic state with lower symmetry. For \BFCA{}, the resistivity of detwinned single crystals shows an appreciable anisotropy in the tetragonal phase up to a temperature well above \Ts{} ($\sim$ 170 K for $x=0$ and 0.04 and $\sim$ 200 K for $x=0.02$). Spectroscopic probes, such as ARPES \cite{Yi2011}, X-ray linear dichroism \cite{Kim2011} and optical spectroscopy \cite{Dusza2011}, also observed the anisotropy at temperatures above \Ts{}. A question is whether these experimental results point toward a nematic electronic state. In our previous study of annealed \BFA{} with \Ts{}=143 K, the optical conductivity spectra showed no appreciable anisotropy at 150 K \cite{Nakajima2011}. The applied unidirectional pressure was $\sim$ 50 MPa larger than that used by Chu \textit{et al.}\ \cite{Chu2010}, but the spectrum at 150 K was nearly isotropic, indicating that the applied pressure did not enhance the anisotropy or extend the anisotropic region above \Ts{}. Although the non-equivalence between 3d$_{xz}$ and 3d$_{yz}$ orbitals is observed above \Ts{} in the spectroscopic measurements on \BFA{} \cite{Yi2011,Kim2011}, the observed temperature is around 150 K, which is just above \Ts{} of the present high-quality crystal. In fact, these measurements were done on as-grown crystals, which show sizable resistivity anisotropy at temperatures well above \Ts{}. In the light of our experimental results \cite{Ishida}, as-grown crystals contain fairly large amount of impurities and/or lattice defects, which would broaden the transition or make \Ts{} distributed over some temperature range. Thus, it is likely that the observed anisotropy above \Ts{} is extrinsic in origin and there is no need to explain in terms of nematicity.



In summary, we have revealed the highly polarizable anisotropic electronic state realized in the AFO phase from the in-plane optical spectra of detwinned \BFCA{}. Upon Co doping, the anisotropy in the finite-energy conductivity decreases as a result of the weakening of the order, in contrast to the resistivity anisotropy becoming sizable. The resistivity anisotropy results from the doped Co atom anisotropically polarizing its surroundings and working as an anisotropic scatterer in the AFO phase. The anomalous anisotropy in the specific optical phonon mode at 257 \cm{} also indicates that dynamical electronic polarization is induced by this lattice vibration. These results are suggestive of involvement of the orbital degree of freedom in the formation of the AFO order.


\begin{acknowledgments}

M.N. and S.I. thank the Japan Society for the Promotion of Science (JSPS) for the financial support. This work was supported by the Japan-China-Korea A3 Foresight Program and Grant-in-Aid for JSPS Fellows from JSPS, Grant-in-Aid for Scientific Research from JSPS and the Ministry of Education, Culture, Sports, Science, and Technology, Japan, and Strategic International Collaborative Research Program (SICORP) from Japan Science and Technology Agency.

\end{acknowledgments}

\end{document}